\documentclass[aps,prl,amsfonts,amsmath,amssymb,reprint,twocolumn,superscriptaddress,showpacs,a4paper]{revtex4-1}

\usepackage[squaren]{SIunits}
\usepackage[usenames,dvipsnames]{color}
\usepackage[pdftex]{graphicx}
\usepackage{bm}
\usepackage{hyperref}
\usepackage{slashed}

\usepackage[normalem]{ulem}

\definecolor{mygreen}{rgb}{0,0.5,0} 
\definecolor{myblue}{rgb}{0,0,0.75} 
\definecolor{mymagenta}{cmyk}{0,1,0,0.12}

\newcommand{\rb}{^{87}\mathrm{Rb}}
\newcommand{\var}[1]{\mathrm{var}(#1)}
\newcommand{\cov}[1]{{\mathrm{cov}(#1})}

\newcommand{\ket}[1]{\ensuremath{\lvert#1\rangle}}
\newcommand{\ave}[1]{\ensuremath{\langle#1\rangle}}
\newcommand{\norm}[1]{\ensuremath{\lvert#1\rvert}}
\newcommand{\NA}{N_{\rm A}}

\newcommand{\NL}{N_{\rm L}}

\newcommand{\Tx}{\hat{T}_{\rm x}}
\newcommand{\TxEff}{T_{\rm x}}
\newcommand{\TyEff}{T_{\rm y}}
\newcommand{\Ty}{\hat{T}_{\rm y}}
\newcommand{\Txy}{\hat{T}_{\rm x,y}}
\newcommand{\Fz}{\hat{F}_{\rm z}}

\newcommand{\fz}{\hat{f}_{\rm z}}
\newcommand{\tx}{\hat{t}_{\rm x}}
\newcommand{\ty}{\hat{t}_{\rm y}}
\newcommand{\dFz}{\var{\FT}}
\newcommand{\dFzOld}{\var{\Fz}}

\newcommand{\Sx}{\hat{S}_{\rm x}}
\newcommand{\Sy}{\hat{S}_{\rm y}}
\newcommand{\Sz}{\hat{S}_{\rm z}}

\newcommand{\supin}{^{({\rm in})}}
\newcommand{\supout}{^{({\rm out})}}

\newcommand{\ph}{\hat{\phi}}
\newcommand{\phA}{\phi_1}
\newcommand{\phB}{\phi_2}

\newcommand{\phAOC}{\hat{\phi}_{\rm AOC}}

\newcommand{\phN}{\ph^{(0)}}

\newcommand{\dd}{{\rm D}_2}
\newcommand{\kA}{\kappa_1}
\newcommand{\kB}{\kappa_2}
\newcommand{\kAB}{\kappa_{1,2}}

\newcommand{\HH}{\hat{H}_{\rm eff}}

\newcommand{\FT}{\declareslashed{}{\text{-}}{0.11}{0}{T}\hat{\slashed{T}}}

\begin{document}

\title{Magnetic sensitivity beyond the projection noise limit by spin squeezing}

% repeat the \author ..  \affiliation  etc.  as needed
% author.  Explanatory text should go in the []'s, actual e-mail

% \affiliation command.  The \affiliation command should follow the
\author{R.J.~Sewell}
\email[]{robert.sewell@icfo.es}
\homepage[]{http://www.mitchellgroup.icfo.es/}
\affiliation{ICFO-Institut de Ciencies Fotoniques, Av. Carl Friedrich Gauss, 3, 08860 Castelldefels, Barcelona, Spain.}
	
\author{M.~Koschorreck}
\affiliation{ICFO-Institut de Ciencies Fotoniques, Av. Carl Friedrich Gauss, 3, 08860 Castelldefels, Barcelona, Spain.}
\affiliation{Cavendish Laboratory, University of Cambridge, JJ Thompson Avenue, Cambridge CB3 0HE, United Kingdom.}

\author{M.~Napolitano}
\affiliation{ICFO-Institut de Ciencies Fotoniques, Av. Carl Friedrich Gauss, 3, 08860 Castelldefels, Barcelona, Spain.}
	
\author{B.~Dubost}
\affiliation{ICFO-Institut de Ciencies Fotoniques, Av. Carl Friedrich Gauss, 3, 08860 Castelldefels, Barcelona, Spain.}
\affiliation{Univ. Paris Diderot, Sorbonne Paris Cit\'{e}, Laboratoire Mat\'{e}riaux et Ph\'{e}nom\`{e}nes Quantiques, UMR 7162, B\^{a}t. Condorcet, 75205 Paris Cedex 13, France}	
	
\author{N.~Behbood}
\affiliation{ICFO-Institut de Ciencies Fotoniques, Av. Carl Friedrich Gauss, 3, 08860 Castelldefels, Barcelona, Spain.}
	
\author{M.W.~Mitchell}
\affiliation{ICFO-Institut de Ciencies Fotoniques, Av. Carl Friedrich Gauss, 3, 08860 Castelldefels, Barcelona, Spain.}
\affiliation{ICREA-Instituci\'{o} Catalana de Recerca i Estudis Avan\c{c}ats, 08015 Barcelona, Spain}

\date{\today}

\begin{abstract}
We report the generation of spin squeezing and entanglement in a magnetically-sensitive atomic ensemble, and entanglement-enhanced field measurements with this system.  
A maximal $m_f=\pm1$ Raman coherence is prepared in an ensemble of $8.5\times10^5$ laser-cooled $\rb$ atoms in the $f=1$ hyperfine ground state, and the collective spin is squeezed by synthesized optical quantum non-demolition measurement.  
This prepares a state with large spin alignment and noise below the projection-noise level in a mixed alignment-orientation variable.  3.2\,dB of noise reduction is observed and 2.0\,dB of squeezing by the Wineland criterion, implying both entanglement and metrological advantage.  
Enhanced sensitivity is demonstrated in field measurements using alignment-to-orientation conversion.
\end{abstract}

% \keywords{}

\maketitle

Spin-squeezing of atomic ensembles~\cite{Wineland1992,Kitagawa1993} via quantum non-demolition (QND) measurement~\cite{Kuzmich1998} is of both fundamental and practical interest. 
Through spin-squeezing inequalities, collective (and thus macroscopic) observables imply underlying microscopic entanglement~\cite{Sorensen2001b,Toth2007}, the basic resource for applications in quantum information~\cite{Lvovsky2009,*Cviklinski2007,*Dubost2012} and quantum simulation~\cite{Eckert2008,*Toth2010}. 
In quantum metrology~\cite{Giovannetti2006,*Giovannetti2011}, spin squeezing promises to improve atomic sensors~\cite{Huelga1997,*Andre2004,Auzinsh2004,Kominis2008,Vasilakis2011}, including in the spin-exchange-limited regime~\cite{Sheng2012}.
In particular, in optical magnetometers QND measurement has been experimentally demonstrated to increase bandwidth~\cite{Shah2010,*Wasilewski2010a} and improve sensitivity~\cite{Sheng2012} in state-of-the-art devices.

While measurement-induced squeezing of real~\cite{Takano2009} and effective~\cite{Appel2009,*Schleier-Smith2010a,*Chen2011} spin-1/2 systems has been demonstrated, it has proven challenging in the large-spin ($f>1/2$) systems used in atomic magnetometry~\cite{Geremia2008}. 
Here we generate squeezing in a spin-1 ensemble using synthesized QND measurements \cite{Koschorreck2010b}, a technique applicable also to larger spin.  
The observed squeezing implies both entanglement among the spin-1 atoms~\cite{Sorensen2001b} and metrological advantage by the Wineland criterion \cite{Wineland1992}.  
As we demonstrate, it measurably improves sensitivity in field measurements by alignment-to-orientation conversion (AOC)~\cite{Budker2000,*Budker2000a,*Budker2002,Pustelny2008}.

In contrast to work based on precession of spin \emph{orientation} ${\bf f}$  \cite{Geremia2008}, our QND and magnetometry strategies employ spin \emph{alignment} ${\bf t}$, i.e., Raman or $\Delta m=2$ coherence, which naturally arises in optical interaction with large spins \cite{Budker2000,*Budker2000a,*Budker2002}.  
We prepare a large alignment by optically pumping the  $^{87}$Rb ensemble into a superposition of the $\ket{f=1,m=\pm1}$ states.  Alignment precesses in response to magnetic fields and can be measured optically.  
Projection-noise-level measurements, however, require probing strong enough to convert alignment to orientation via stimulated Raman transitions \cite{Geremia2006,*Echaniz2008}.  
This AOC during probing results in measurement of a mixed \emph{alignment-orientation} variable $\FT$ defined below.  
More serious, unchecked AOC couples measurement back-action into the signal of both ${\bf f}$- and ${\bf t}$-based strategies \cite{Koschorreck2010b}, destroying quantum enhancement.  
To achieve squeezing and quantum enhancement in this scenario, we measure $\FT$ with pulse pairs, polarized such that the second pulse ``unwinds'' the AOC produced by the first.  
This allows us to squeeze $\FT$, evade measurement back-action, and observe enhanced sensitivity.

\newcommand{\cE}{\tilde{\cal E}}
\newcommand{\supplus}{^{(+)}}
\newcommand{\supminus}{^{(+)}}
\newcommand{\tSx}{\tilde{S}_{\rm x}}
\newcommand{\tSy}{\tilde{S}_{\rm y}}
\newcommand{\tSz}{\tilde{S}_{\rm z}}

\paragraph{Atomic and optical systems and their interaction.---}
We work with an ensemble of $f=1$ atoms interacting with pulses of near-resonant light propagating along the $z$-axis.  
If ${\bf f}^{(i)}$ is the total spin of the $i$'th atom, then the operators $\Fz\equiv\sum_{i}^{N_{A}}\fz^{(i)}/2$ and $\Txy\equiv\sum_{i}^{N_{A}}\hat{t}^{(i)}_{\rm x,y}$ describe the collective atomic spin orientation and alignment, respectively.  
The operators $\tx \equiv (\hat{f}_{\rm x}^2 - \hat{f}_{\rm y}^2 )/2$ and $\ty \equiv (\hat{f}_{\rm x}\hat{f}_{\rm y} + \hat{f}_{\rm y}\hat{f}_{\rm x})/2$ describe single-atom Raman coherences, i.e., coherences between states with $m_f$ different by 2.  For $f=1$, these obey commutation relations $[\Tx,\Ty]= i\Fz$ and cyclic permutations.  
The light is described by the time-varying Stokes operators ${\bm \tilde{S}}(t)$ defined as  $ \tilde{S}_i \equiv \frac{1}{2} (\cE\supminus_+ ,\cE\supminus_-) \sigma_i (\cE\supplus_+ ,\cE\supplus_-)^T $, where the $ \sigma_i $ are the Pauli matrices and $\cE\supplus_\pm(t)$ are the positive frequency parts of quantized fields for the circular plus/minus polarizations.  
We write pulse-integrated Stokes operators as ${\hat{S}}_i \equiv \int dt \tilde{S}_i(t)$. 
In the experiments $\Sy$ is detected. 

\newcommand{\supone}{^{(1)}}
\newcommand{\suptwo}{^{(2)}}
\newcommand{\supinone}{^{({\rm in},1)}}
\newcommand{\supintwo}{^{({\rm in},2)}}
\newcommand{\supoutone}{^{({\rm out},1)}}
\newcommand{\supouttwo}{^{({\rm out},2)}}
\newcommand{\myomit}[1]{}

As described in references~\cite{Geremia2006,*Echaniz2008}, the light pulses and atoms interact by the effective Hamiltonian
\begin{equation}
	 \HH =  \kA\tSz\Fz + \kB(\tSx\Tx + \tSy\Ty),
	\label{eqn:H_full}
\end{equation}
where $\kAB$ are coupling constants that depend on the beam geometry, excited-state linewidth, laser detuning, and the hyperfine structure of the atom.  The resulting evolution is described by 
\begin{eqnarray}
	\label{eqn:SyEv}
	\tSy\supout&=&\tSy\supin + \kA \tSx\supin \Fz - \kB \tSz\supin \Tx 
\end{eqnarray}
where the superscripts $\supin,\supout$ indicate operators at the input and the output of the ensemble, and by
\begin{eqnarray}
	\label{eqn:TyFzEv}
	\frac{d}{dt} \left( \begin{array}{c}\Tx \\ \Ty\\ \Fz \end{array} \right) &=& \left( \begin{array}{ccc} 0 & -\kA \tSz & \kB \tSy \\
	\kA \tSz &  0 & -\kB \tSx  \\ -\kB\tSy &  \kB \tSx  & 0 \end{array} \right) \left( \begin{array}{c} \Tx \\ \Ty\\ \Fz \end{array} \right).
\end{eqnarray}
In all scenarios of interest $|\ave{\Tx}| \approx N/2 \gg |\ave{\Ty}|,|\ave{\Fz}|$.  

The $\kA$ term in Eq.~\ref{eqn:H_full} describes paramagnetic Faraday rotation: it rotates the light polarization in the $\Sx,\Sy$ plane by an angle $\propto \Fz$. 
Acting alone, this describes a QND measurement of $\Fz$, i.e., with no back-action on $\Fz$. The $\kB$ term, in contrast, contributes an $\Sz\rightarrow \Sy$ optical rotation $\propto T_x$ with back-action described by Eq. (\ref{eqn:TyFzEv}).  
We use the dynamics of Eqs.~\ref{eqn:SyEv},~\ref{eqn:TyFzEv} in three ways:

1) We make a direct dispersive measurement of the Raman coherence $\Tx$, i.e., the collective alignment, by probing with an $\Sz$-polarized pulse $\ave{\Sz\supin} = N/2$.  
This measurement is made using an auxiliary probe beam, with coupling constants $\kA^{({\rm aux})},\kB^{({\rm aux})}$.
Integrating Eq.  \ref{eqn:SyEv}  and dropping the tiny $\kA^{({\rm aux})} \Sx\Fz$ term gives
 \begin{equation}
	\Sy\supout =  \Sy\supin - \kB^{({\rm aux})} \Sz\supin \Tx\supin\ .
\end{equation}
 
% This gives a direct dispersive measurement of the Raman coherence $\Tx$, i.e., the collective alignment.
 
2) We make an alignment-to-orientation conversion measurement using a single, $\Sx$-polarized pulse.  
Integrating Eqs (\ref{eqn:SyEv}), (\ref{eqn:TyFzEv}), and keeping terms to second order in $\Sx$, we find
\begin{eqnarray}
 \label{eqn:AOCSignal}
	\Sy\supout &=& \Sy\supin+\kA\Sx\supin \Fz\supin+\frac{\kA \kB}{2}[\Sx\supin]^2\Ty\supin \nonumber \\
		&=& \Sy\supin+\frac{\kA\ave{\Sx}}{\cos\theta}\FT\supin.
\end{eqnarray}
This describes a measurement, with optical read-out noise $\Sy\supin$, of a mixed alignment-orientation variable $\FT \equiv \Fz \cos \theta + \Ty \sin \theta$, where $\tan \theta =  \kB \ave{\Sx} /2$ and $\ave{\Sx}=\NL/2$. 
During the measurement the atomic variables experience a rotation in the $\Ty$-$\Fz$ plane due to the $\kB$ term in Eq.~\ref{eqn:H_full}.

3) We synthesize a QND measurement free of $\kB$-induced rotations using pulses of alternating polarization, as described in Ref.~\cite{Koschorreck2010b}.
Two successive pulses, polarized $\ave{\Sx^{(\rm in,1)}}=-\ave{\Sx^{(\rm in, 2)}}=\NL/2$, give signals $\Sy^{\rm (out,1)}$ and $\Sy^{\rm (out,2)}$, respectively. 
The atomic evolution during the second pulse is the time-reverse of the evolution during the first.  
As a result, there is no net  $\Ty \rightleftarrows \Fz$ rotation, protecting the measured variable from probe-induced decoherence.  
Furthermore, the differential signal, found again by integration of Eqs (\ref{eqn:SyEv}), (\ref{eqn:TyFzEv}), is 
 \begin{eqnarray}
	\delta \Sy\supout & \equiv & \Sy\supouttwo - \Sy\supoutone \nonumber \\ 
	&=& \Sy\supintwo - \Sy\supinone +2\frac{\kA\ave{\Sx}}{\cos\theta}\FT\supin
\end{eqnarray}
This measures the same variable $\FT$ as in the single-pulse case.  
Because of this, we can use the synthesized QND measurement to prepare a squeezed state with reduced noise in the same atomic variable detected by the AOC measurement.

\begin{figure}[t]
	\includegraphics[width=\columnwidth]{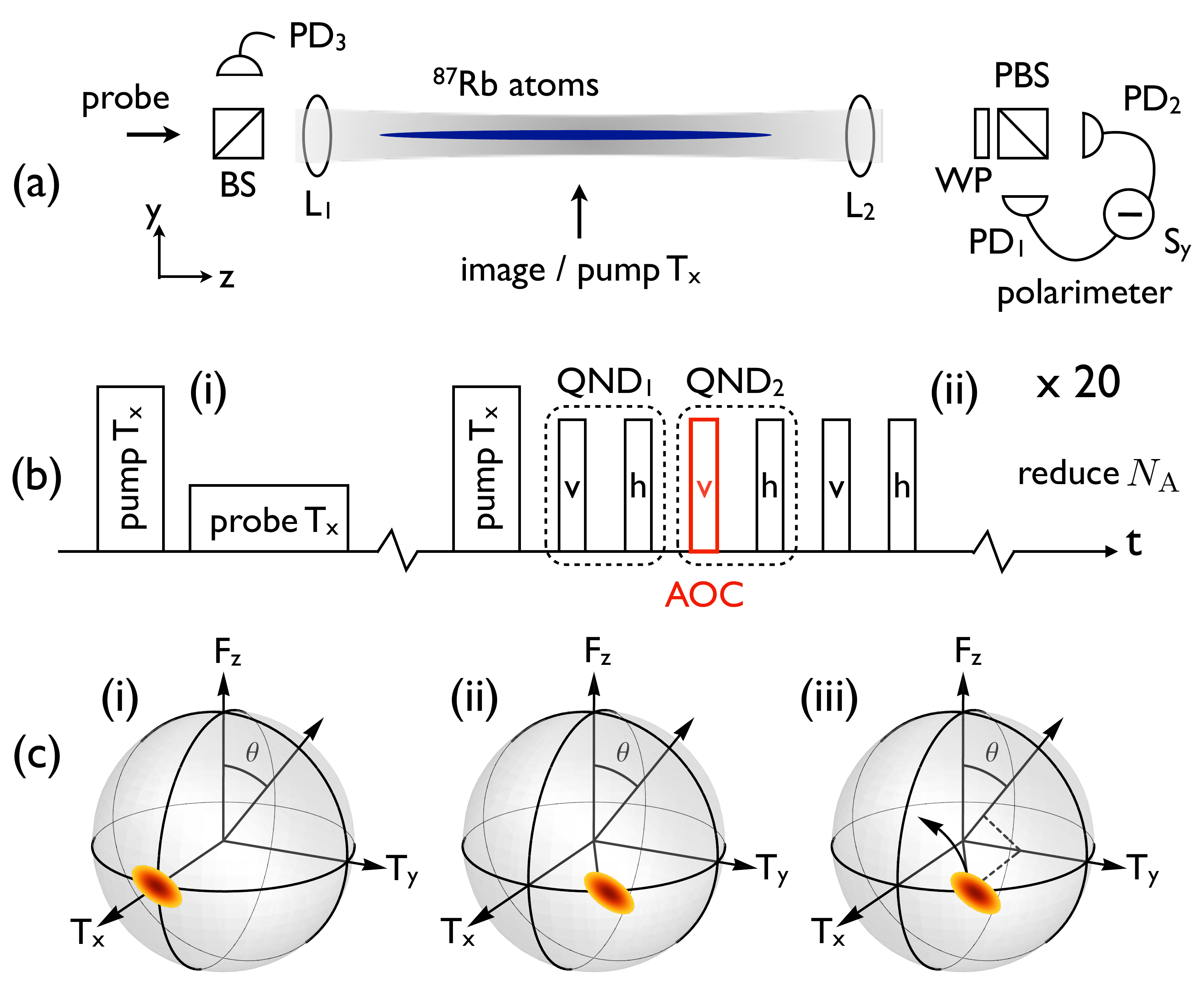}%
	\caption{
	{\bf (a)} Experimental geometry.  
	PD: photodiode; L: lens; WP: waveplate; BS: beam-splitter; PBS: polarizing beam-splitter.  
	{\bf (b)} Measurement pulse sequence for {\bf (i)} dispersive measurement of Raman coherence and {\bf (ii)} QND measurement and alignment-to-orientation conversion.  
	{\bf (c)}  Entanglement-enhanced field measurement:  {\bf (i)} a two-pulse QND measurement prepares a $\TxEff$-aligned state with reduced uncertainty in an alignment-orientation variable with a mixing angle $\theta$ that {\bf (ii)} rotates into $\TyEff$ due to Zeeman shifts and {\bf (iii)} is then read out by a single pulse, giving an integrated signal proportional to the Zeeman shift.  
	See text for details.
	\label{fig:setup}
	}
\end{figure}

\paragraph{Experiments.---}
We work with an ensemble of up to $8.5\times10^5$ laser cooled $\rb$ atoms in the $f=1$ ground state, held in a weakly focused (\unit{52}{\micro\meter} beam waist) single beam optical dipole trap, as illustrated in Fig.~\ref{fig:setup}(a) and described in detail in Ref.~\cite{Kubasik2009}.  
The fitted size of the cloud is \unit{3}{\milli\meter} by \unit{20}{\micro\meter} full width at half maximum, (FWHM) indicating an atomic density at the center of approximately \unit{5\times10^{11}}{atoms\per\centi\metre^3}.
We probe the atoms with \unit{\micro\second} pulses of near-resonant light on the $\dd$ line, detected by a shot-noise-limited polarimeter.  
The linearly-polarized probe is focused to a waist of \unit{20}{\micro\meter}, matched to the width of the atomic cloud.  
The circularly polarized probe is focused to a waist of \unit{50}{\micro\meter} to ensure a more uniform illumination of the atomic sample.  
This trap geometry produces a large atom-light interaction for light propagating along the trap axis with an effective on-resonance optical depth $d_0=43.5$.  
The experiment achieves projection-noise-limited sensitivity, calibrated against a thermal spin state, with a demonstrated spin read-out noise of $\dFzOld =(515\,{\rm spins})^2$~\cite{Koschorreck2010a}.
We actively cancel homogeneous magnetic fields and field gradients along the length of the trap, leaving a residual bias field $B_z\simeq\unit{100}{\nano\tesla}$ and gradient field components $\partial B_i/\partial z<\unit{200}{\nano\tesla\per\centi\meter}$, which limit the spin coherence time to $\tau_c=\unit{290}{\micro\second}$.

The measurement sequence is illustrated in Fig.~\ref{fig:setup}(b).  
For each measurement we prepare a coherent spin state (CSS) with $\Tx=\NA/2$ via optical pumping and measure the Raman coherence with a train of \unit{\micro\second} pulses of circularly polarized light with $\unit{10^6}{photons\per pulse}$ at a detuning of \unit{190}{\mega\hertz} to the red of the $f=1\rightarrow f'=0$ transition.  
We then re-prepare $\Tx$ and probe the atoms with a train of \unit{2}{\micro\second} long pulses of light every \unit{5}{\micro\second} with $\unit{2\times10^8}{photons\per pulse}$ and alternating $v$- and $h$-polarization at a detuning of \unit{600}{\mega\hertz}.  
These pulses are used in pairs to synthesise the QND measurement, or singly for the alignment-to-orientation conversion measurement.  
We vary the number of atoms, $\NA$, used in the experiment from $3.9\times10^4$ to $8.5\times10^5$ by switching off the optical dipole trap for \unit{100}{\micro\second} after each measurement, which reduces the atom number by $ \sim15\,\% $, and repeating the sequence 20 times per trap loading cycle. 
At the end of each cycle the measurement is repeated without atoms in the trap.  To collect statistics, the entire cycle is repeated 1090 times.  

The dispersive probing of the collective alignment is calibrated against a measurement of the atom number made by absorption imaging, as described in Ref.~\cite{Koschorreck2010a}.
For the circularly polarized probe we measure $\kB^{\rm (aux)}=0.9\times10^{-7}$ radians per spin.
To account for the spatial variation in the coupling between the probe beam and the trapped atoms, we follow Refs.~\cite{Appel2009,*Schleier-Smith2010a,*Chen2011} and define an effective atom number such that the parameteric Faraday rotaion signal is proportional to the total number of atoms, and the expected variance of the measurement variable is $\dFz = \TxEff/2$.  
For our trap and probe geometry $\NA^{\rm eff}=0.9\NA$.
For the linearly polarized probe we measure $\kA=1.47\times10^{-7}$ radians per spin, from which we calculate $\kB=7.54\times10^{-9}$ radians per spin.

% (see Supplementary Information).

\newcommand{\phNsub}{{\rm RO}}
\renewcommand{\phN}{\phi_\phNsub}

\begin{figure}[t]
	\includegraphics[width=\columnwidth]{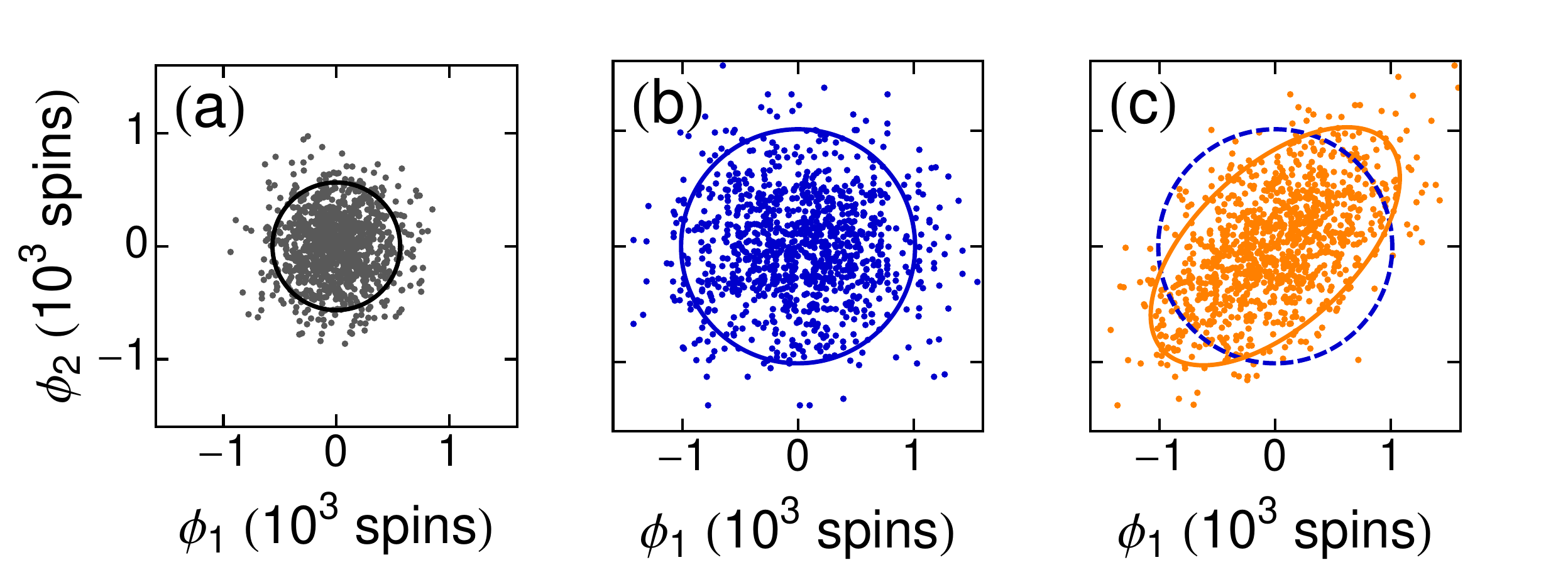}%
	\caption{
	Joint probability distribution of successive QND measurements with {\bf (a)} no atoms in the trap, i.e. read-out noise, {\bf (b)} independent CSS preparations, and {\bf (c)} a single CSS preparation.  
	Solid curves indicate $2\sigma$ radii for gaussian fits.  
	Dashed blue circle in {\bf (c)} reproduces solid circle in {\bf (b)} indicating the standard quantum limit for the input CSS.  
	Note: for presentation purposes, a small mean offset has been subtracted from the data.
	\label{fig:correlation}
	}
\end{figure}

\paragraph{Conditional noise reduction.---}
To study spin squeezing, we synthesize two successive QND measurements of the input coherent spin state, each containing one vertically-- and one horizontally--polarized optical pulse.  
For convenience, we define a normalized measurement variable $\ph\equiv(\cos\theta/\kA\Sx\supin)\Sy\supout$, corresponding to the scaled rotation angle of the input light polarization, so that $\ph=\phN+\FT\supin$, where the read-out noise $\phN$ contains the electronic and light noise contributions to the measurement, quantified by repeating the measurement with no atoms in the trap.
If $\phi^{(n)}$ is the scaled rotation of the $n$'th pulse, the two QND measurements are then $\phA=\ph^{(1)}-\ph^{(2)}$ and $\phB=\ph^{(3)}-\ph^{(4)}$.  The measurement-induced noise reduction is quantified by the conditional variance $\var{\FT|\phA}=\var{\phB-\chi\phA}-\var{\phN}$, where $\chi\equiv\cov{\phA,\phB}/\var{\phA} > 0$ describes the correlation between $\phA$ and $\phB$~\cite{Appel2009}.

Figure~\ref{fig:correlation} shows $\phA$ and $\phB$ correlation plots for $\TxEff=3.4\times10^5$.  
The read-out noise (Fig.~\ref{fig:correlation}(a)) is dominated by light shot-noise: we estimate the technical noise contribution to the read-out at \unit{-19}{dB} compared to the light shot-noise with the number of photons, $\NL=4\times10^8$, used in the QND measurement. 
Measurements of independently prepared $\Tx$-states (Fig.~\ref{fig:correlation}(b)) are uncorrelated, whereas two measurements of the same state (Fig.~\ref{fig:correlation}(c)) are strongly correlated, so that the first measurement can be used to predict the second with an optimal estimator $\chi\phA$ and an uncertainty below the standard quantum limit.

\begin{figure}[t]
	\includegraphics[width=\columnwidth]{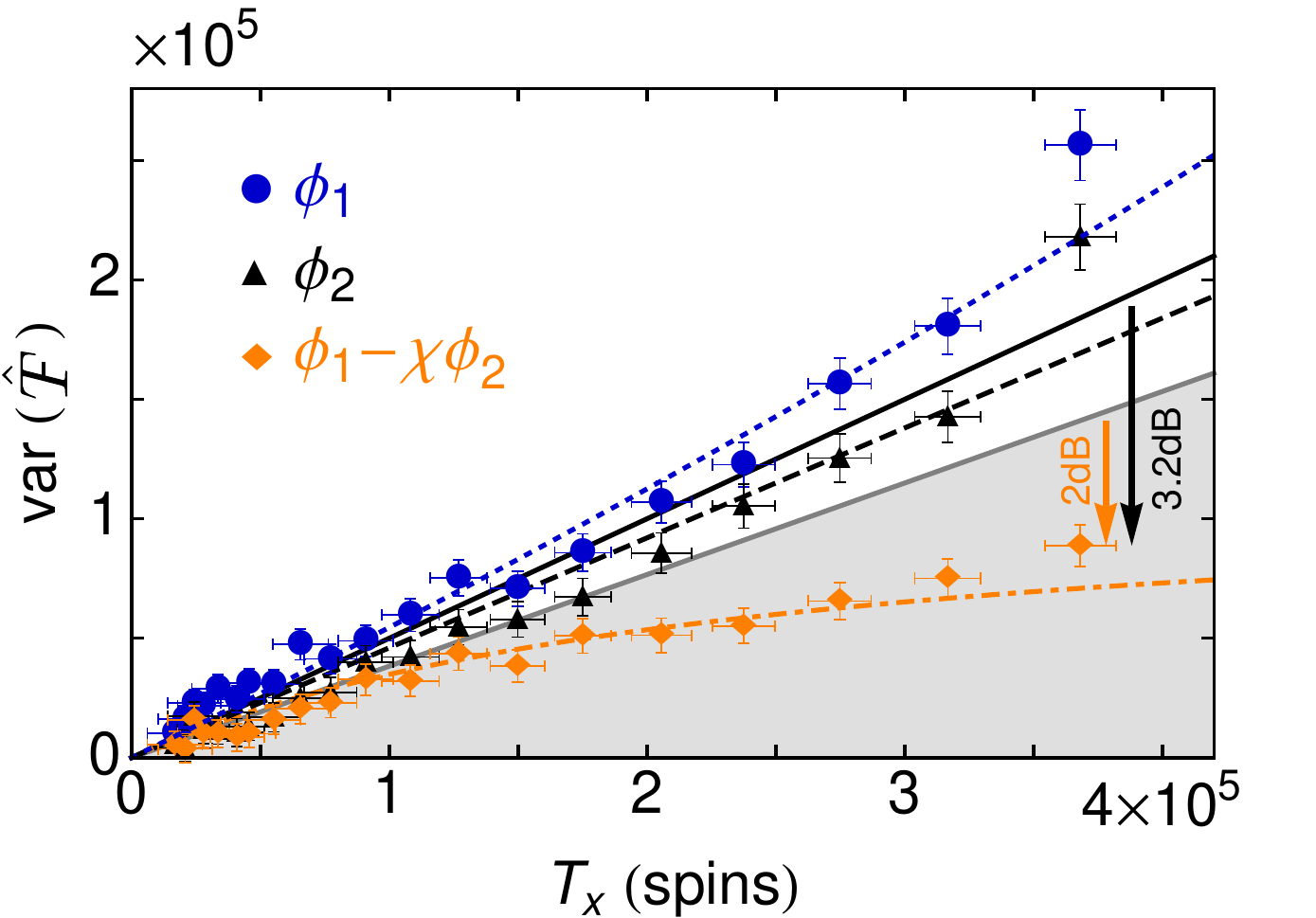}%
	\caption{
	Noise scaling of the QND measurement of the CSS and spin-squeezed state.  
	Shaded area represents region of increased metrological sensitivity due to spin squeezing.  
	Curves are plotted from independently measured experimental parameters unless noted; see text for details.  
	Horizontal and vertical error bars represent $\pm1\sigma$ statistical errors.  
	Read-out-noise has been subtracted from each data set.
	\label{fig:spin_squeezing}
	}
\end{figure}

\paragraph{Spin squeezing \& entanglement.---}
Figure~\ref{fig:spin_squeezing} shows the individual variances of the two QND measurements, $\phA$ and $\phB$ (blue circles and black triangles), as a function of $\TxEff$.  
The collective alignment has an expected variance $\dFz = \TxEff/2$ (solid black line) that scales linearly with $\TxEff$.  
A quadratic fit, $\var{\FT}=a_1\TxEff+a_2\TxEff^2$ (blue dotted line), to the measured data for $\phA$ yields $a_1=0.53(3)$ and $a_2=0.02(1)$, consistent with projection-noise-limited QND measurement.
The atomic technical noise contribution is \unit{7}{dB} below the projection-noise at $\TxEff=3.7\times10^5$. 
Accounting for loss and decoherence~\cite{Echaniz2005}, the expected variance for the second measurement is $\dFz = \TxEff\supout/2=0.46\,\TxEff$ (black dashed line). 
A quadratic fit to the measured data for $\phB$ yields $a_1=0.44(2)$ and $a_1=0.01(1)$.

The measured conditional variance $\var{\FT|\phA}$ (orange diamonds in Figure~\ref{fig:spin_squeezing}), is up to \unit{3.2}{dB} below the projection noise, in agreement with the predicted value (orange dot-dashed line), which is given by $\dFz\supout=\dFz\supin/(1+\zeta)$~\cite{Hammerer2004}, where the signal-to-noise ratio $\zeta\equiv\kA^2\NL\TxEff$ is calculated from independent measurements.

We quantify metrological advantage by the Wineland criterion~\cite{Wineland1992}, which accounts for both the noise and the coherence of the post-measurement state: if $\xi_{\rm m}^2 \equiv \dFz\supout\TxEff\supin/2\norm{\TxEff\supout}^2$, then $\xi_{\rm m}^2<1$ indicates metrological advantage.  
The post-measurement spin alignment is $\TxEff\supout=(1-\eta_{\rm sc})(1-\eta_{\rm dep})\TxEff\supin$, where $\eta_{\rm sc}=0.093$ and $\eta_{\rm dep}=0.034$ are independently-measured depolarizations due to probe scattering and field inhomogeneities, respectively.  
The contribution from $\eta_{\rm dep}$ could be recovered by spin-echo techniques.  
$|\Tx\supout|^2$ is indicated by a gray line in Fig.~\ref{fig:spin_squeezing}; a conditional variance below this line (shaded region) gives metrological advantage.  
For $\TxEff=3.7\times 10^{5}$ we find $\xi_{\rm m}^2=0.63$, or \unit{2.0}{dB} of metrologically useful spin squeezing.  
We note that for large-spin atoms, it is possible to squeeze the internal state of individual atoms without squeezing the collective state~\cite{Fernholz2008,*Rochester2012}.  
In our experiment, the observed noise reduction of \unit{3.2}{dB} and alignment of $\TxEff\supout=0.88\,\TxEff\supin$ are sufficient to imply entanglement among the spin-1 atoms~\cite{Sorensen2001b}.

\newcommand{\phAOCN}{\phi_{\rm AOC,RO}}

\begin{figure}[t]
	\includegraphics[width=\columnwidth]{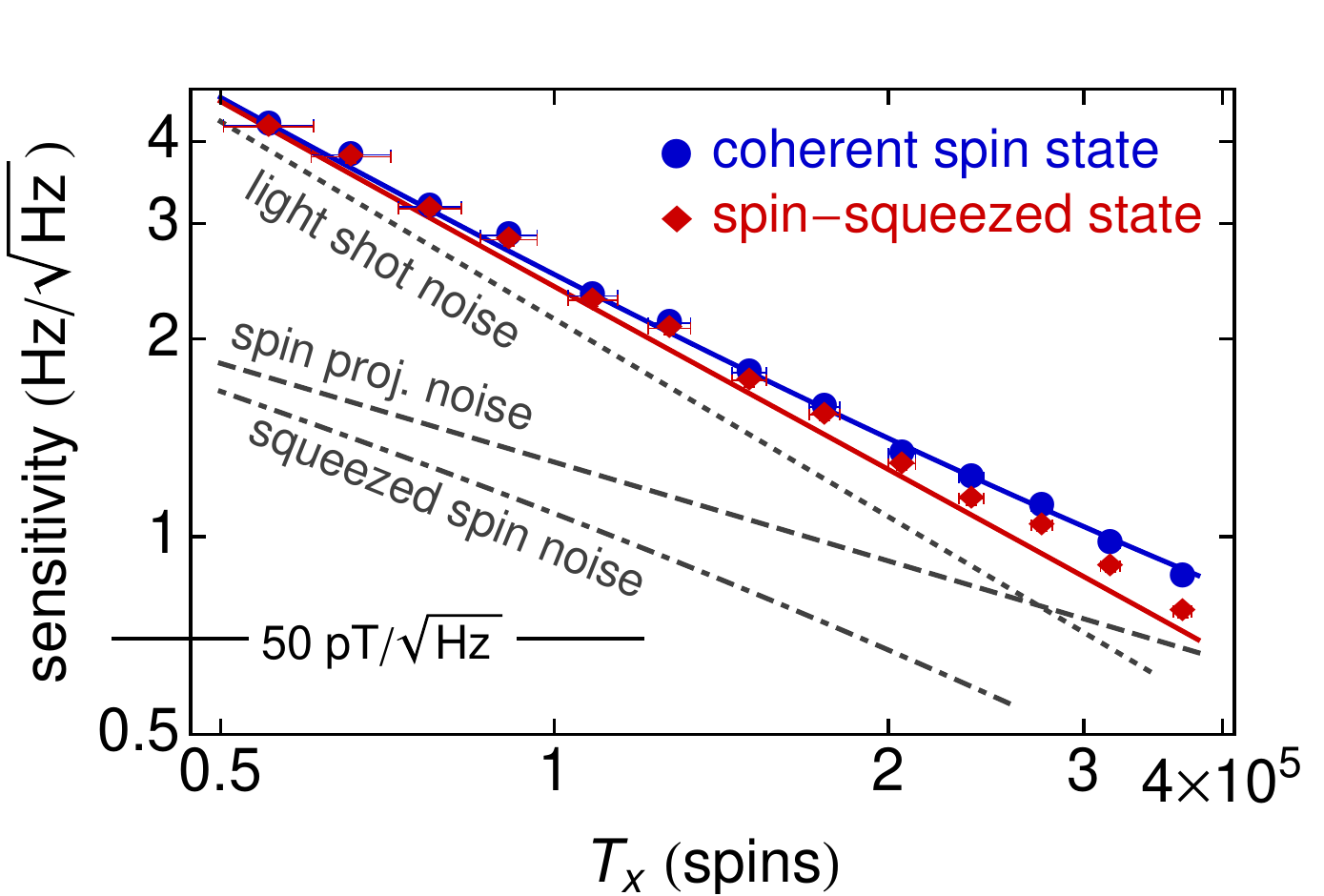}%
	\caption{
	Log-log plot of alignment-to-orientation measurement sensitivity with an input coherent spin state (blue circles) and spin-squeezed state (red diamonds).  
	Curves are plotted from independently measured experimental parameters.  
	Solid curves are the predicted measurement sensitivity with an input CSS (blue), and spin-squeezed state (red) with atomic projection noise reduced by a factor $1/(1+\zeta)$.  
	Broken curves represent the light shot-noise (dotted), atomic projection-noise (dashed) and spin-squeezed projection-noise (dot-dashed) contributions to the sensitivity.  
	Horizontal and vertical error bars represent $\pm1\sigma$ statistical errors.
	\label{fig:sensitivity}
	}
\end{figure}

\paragraph{Squeezing-enhanced field measurement.---}
We use the spin squeezed state to enhance the sensitivity of a field measurement using aligment-to-orientation conversion.  
We employ a Ramsey sequence (illustrated in Fig.~\ref{fig:setup}(c)): 
First, two pulses are used as above to synthesize a QND measurement with result $\phA$.  
This prepares a $\Tx$-aligned, $\FT$-squeezed state.  
Next, the state is allowed to evolve for a time $T=\unit{5}{\micro\second}$, giving a measurement bandwidth of $T^{-1}=\unit{200}{\kilo\hertz}$, during which time residual Zeeman shifts (magnetic or optical) cause a state rotation $\Tx\rightarrow \Ty$.  
Finally, a third, single vertically--polarized pulse is used for an alignment-to-orientation measurement, giving an optical rotation $\phAOC$.  
As described in Eq.~\ref{eqn:AOCSignal}, this measurement gives a signal $\ave{\phAOC}=\tfrac{1}{2}\kB\ave{\Sx\supin}\ave{\Ty}$ and noise $\var{\phAOC}=\var{\phAOCN}+\var{\FT}$.  
The metrological advantage is quantified by the conditional variance $\var{\phAOC|\phA} = \var{\phAOC-\chi\phA}$,  $\chi\equiv\cov{\phA,\phAOC}/\var{\phA}$.  
We observe an average signal $\ave{\Ty}=0.033(2)\TxEff$, corresponding to an energy shift of $\unit{2.9}{\kilo\hertz}$ between the $m=\pm 1$ Zeeman states.

In Fig.~\ref{fig:sensitivity} we plot the measurement sensitivity $\delta E/h=\Delta\FT/(\kB\ave{\Sx}\ave{\Tx}\sqrt{T})$ as a function of $\TxEff$ for an input coherent spin state (blue circles) and spin-squeezed state (red diamonds).  
We emphasise that read-out noise has not been subtracted from these data.  
As expected, the sensitivity advantage due to spin squeezing increases with $\TxEff$, both because the degree of squeezing increases and because atomic projection noise is a larger fraction of the measurement noise.  
We observe up to $5\,\sigma$ gain in measurement sensitivity at large $\TxEff$ due to the spin-squeezing, with a maximum of $11\pm2\%$ improvement.

Spin squeezing is expected to improve short-term sensitivity, and thus the measurement bandwidth, when the spin coherence time exceeds the measurement time~\cite{Huelga1997,Auzinsh2004,Shah2010}, as demonstrated here.
In addition, long-term sensitivity can also be improved in high-density, highly polarized ensembles due to the suppression of spin-relaxation noise~\cite{Kominis2008,Vasilakis2011}.
It is therefore interesting to compare our results to the best reported optical magnetometers.  
Our energy sensitivity corresponds to a  field sensitivity of $\delta B=\unit{105}{\femto\tesla\per\sqrt{\centi\meter^{3} \hertz}}$ in a single-shot measurement with a bandwidth of \unit{200}{\kilo\hertz} and a measurement volume $V=\unit{3.7\times10^{-6}}{\centi\meter^3}$. 
For comparison, the best chip-scale vapor cell magnetometers report $\delta B\sim \unit{10}{\femto\tesla\per\sqrt{\centi\meter^{3} \hertz}}$ with volumes $\sim\unit{10^{-3}}{\centi\meter^{3}}$ and measurement bandwidths of \unit{1}{\kilo\hertz} or less~\cite{Smullin2009,Griffith2010}.   
Improving $\tau_c$ from \unit{290}{\micro\second} to \unit{5}{\milli\second}, e.g. by active~\cite{Seltzer2004,*Smith2011} or passive~\cite{Kominis2003,*Kornack2007} field gradient control, would allow us to extend inter-measurement precession time from \unit{5}{\micro\second} to \unit{500}{\micro\second} and thereby boost projection-noise-limited sensitivity to $\delta B \approx \unit{10}{\femto\tesla\per\sqrt{\centi\meter^{3}\hertz}}$ with \unit{2}{\kilo\hertz} bandwidth.

In summary, we have demonstrated squeezing of spin orientation by quantum non-demolition measurement in a spin-aligned atomic ensemble with up to $8.5\times10^5$ laser-cooled $\rb$ atoms in the $f=1$ hyperfine ground state.  
We observe \unit{3.2}{dB} of quantum noise reduction and \unit{2.0}{dB} of metrologically-relevant spin squeezing, implying entanglement among the spin-1 atoms, consistent with theory and limited by the optical depth and dephasing due to residual magnetic field inhomogeneity.  
We use the spin-squeezed state to make an entanglement-enhanced alignment-to-orientation conversion measurement of the Zeeman shift of the $m_f=\pm1$ sublevels, with a direct gain in measurement sensitivity of $11\%$ without noise subtraction.
The techniques used here: stroboscopic probing and quantum non-demolition measurement, may also enable sub-projection-noise sensitivity in magnetometry with dense, spin-exchange-limited atomic vapors~\cite{Sheng2012}, as well as applications in quantum state manipulation~\cite{Norris2012}, quantum information~\cite{Lvovsky2009,*Cviklinski2007,*Dubost2012}, and quantum simulation~\cite{Eckert2008,*Toth2010}.
% In addition to improving the sensitivity of optical magnetometry by alignment-to-orientation conversion~\cite{Budker2000,*Budker2000a}, our techniques may be advantageous for applications in quantum state manipulation~\cite{Norris2012}, quantum information~\cite{Lvovsky2009,*Cviklinski2007,*Dubost2012}, and quantum simulation~\cite{Eckert2008,*Toth2010}.

% Specify following sections are appendices. Use \appendix if there
% \appendix

\begin{acknowledgments}
This work was supported by the Spanish MINECO under the project MAGO (Ref. FIS2011-23520), Fundacio Cellex Barcelona, and the European Research Council under the project AQUMET.
\end{acknowledgments}


% See Supplemental Material at [URL will be inserted by publisher] for [give brief description of material]

% \bibliography{SpinSqueezingReferences}


\end{document}